\documentclass[10pt]{article}
\usepackage{graphicx}

\usepackage{epstopdf}

\usepackage{verbatim}

\begin{document}

\begin{center}

\Large{\bf A note on linearized stability of Schwarzschild thin-shell \\ wormholes with variable equations of state}

\vskip 1truecm

\large{Victor Varela\footnote{Email: victor.varela.abdn@gmail.com}\\

9616 Castle Ridge Circle, Highlands Ranch, CO 80129, USA\footnote{Previous institutional address: Institute of Mathematics, University of Aberdeen, King's College, Aberdeen AB24 3UE, UK.}}

\vskip 1truecm

March 20, 2015

\end{center}\emph{}

\begin{abstract}
We discuss how the assumption of variable equation of state (EoS) allows the elimination of the instability at equilibrium throat radius $a_0=3M$ featured by previous Schwarzschild thin-shell wormhole models. Unobstructed stability regions are found for three choices of variable EoS. Two of these EoS entail linear stability at every equilibrium radius. Particularly, the thin-shell remains stable as $a_0$ approaches the Schwarzschild radius $2M$. A perturbative analysis of the wormhole equation of motion is carried out in the case of variable Chaplygin EoS. The squared proper angular frequency $\omega_0^2$ of small throat oscillations is linked with the second derivative of the thin-shell potential. In various situations $\omega_0^2$ remains positive and bounded in the limit $a_0\rightarrow 2M$.\\
\\
PACS indexing codes: 04.20.-q, 04.20.Jb
\end{abstract}

\vskip 1truecm

\section{Introduction}
In 1966 Israel \cite{israel} presented a powerful formalism to describe spacetime junctions. Continuity of the spacetime metric is generally required on the hypersurfaces defined by these junctions. Discontinuities of the extrinsic curvature on these hypersurfaces lead to concentrations of energy-momentum called thin shells. The dynamical framework for these zero-thickness objects follows from the Einstein equations. These sources are modeled as two-dimensional fluids characterized by surface energy density and surface pressures. Spherically-symmetric thin shells involve only one surface pressure, and their dynamical description is greatly simplified.

Thin shells are used in different ways within general relativity. Kijowski et al. \cite{kmm} as well as Krisch and Glass \cite{kg} listed research lines in gravitational theory involving the use of these objects. And the number of applications of Israel's thin-shell formalism keeps growing. (See, for example, the very recent contributions by Mazharimousavi et al. \cite{mha} and Pereira et al. \cite{pcr}.)

There is no clear-cut separation between thin-shell and continuum models in all circumstances. In some cases the dynamics of a continuous source naturally leads to concentrations of energy-momentum which are best described as thin shells. An example is the emergence of a thin shell during the collapse of a massless scalar field reported by Beauchesne and Edery \cite{be}.

The question arises whether gravitational source construction based on thin shells rather thick shells provides physically meaningful results. Certainly, the expected dynamical complexity of thick shells suggests the insufficiency of thin-shell models. On the other hand, recent research work indicates that these two approaches complement each other in a number of ways.

Complementary aspects of thick shells and thin shells appear in the sequential development of analytical gravastar models. Mazur and Mottola \cite{mm} argued for the thermodynamic stability of a gravastar model involving a (relatively thin) finite-thickness shell bounded by two thin shells. Visser and Wiltshire \cite{vw} considered the dynamical stability of a thin-shell gravastar retaining most of the original features of Mazur and Mottola's model. Chirenti and Rezzola \cite{cr} studied the stability of a spherical thick-shell gravastar model submitted to axial perturbations. The analysis of non-radial gravastar perturbations was extended by Pani et al. \cite{pbccn}, who combined standard perturbation theory with Israel's formalism to describe polar perturbations of non-rotating, thin-shell gravastar models.

The interplay of thick shells and thin shells emerges in other dynamical contexts as well. Garfinkle and Gregory \cite{gg} expanded the equations of a thick gravitating wall in powers of the thickness of the wall. They found that the zeroth-order equations reproduce the thin-wall approximation for domain walls provided by the Israel formalism. Also, they used the first-order thick-wall equations to modify thin-wall equations and discuss the motion of domain walls. Khakshournia and Mansouri \cite{km} determined that the zeroth-order approximation of the equations of motion of a spherical dustlike thick shell coincide with equations arising from the thin-shell formalism. They also found that the effect of thickness is to speed up the collapse of the shell. More recently, Drobov and Tegai \cite{dt} added anisotropy to a thick wall of fluid and obtained the appropriate thin-shell limit.

Garcia et al. (GLV) \cite{glv} have summarized the status of the energy conditions of standard general relativity in connection with the theoretical construction of wormholes. These hypothetical tunnels in spacetime are supported by "exotic matter" which violates all the pointwise and all the averaged energy conditions. To minimize the extent of this problem, researchers have considered truncated Morris-Thorne models featuring a cut-off of the stress-energy, and a junction interface. (See, for example, \cite{lobo,kuh2}.) This junction constitutes a thin-shell or a boundary surface depending on the glued metrics. A more radical approach is based on the cut-and-paste procedure proposed by Visser \cite{visser}, who joined two identical sections of Schwarzschild's spacetime excluding horizons and black hole interiors. A a result, all the exotic matter concentrates at the arising thin-shell throat.

An evolving, truncated wormhole model with boundary surfaces separating the fluid-filled regions from vacuum, and which totally excludes thin shells could be interpreted as a pure thick shell. It would be interesting to implement Garfinkle and Gregory's approach in this context as well, and expand the dynamical equations of this object in powers of the thickness of the throat. The question arises whether the zeroth-order equations would describe the Schwarzschild thin-shell wormhole. Such a result would provide a fresh justification of Visser's construction.

The derivation of thin-shell wormhole dynamics as the zero-thickness limit of a pure thick-shell model is an open problem. Its eventual solution would support the complementarity of thin-shell and thick-shell approaches to this type of sources. The construction of evolving, pure thick-shell wormholes; the dynamical interpretation of their zero-thickness limit; and their possible connections with thin-shell models are beyond the scope of the present paper.

In the spherically symmetric case Visser's original construction provides objects with relatively simple dynamics whose energy-momentum contents ultimately originate from topological identification. This is a distinguished feature in gravitational physics. The resulting thin-shell wormhole models have only one degree of freedom (throat radius), and are compatible with a variety of dynamics through different choices of one EoS linking surface pressure $p$ with surface energy density $\sigma$. The present author believes that, despite the missing link with thick-shell dynamics, the topological character of thin-shell wormholes, their highly simplified equations of motion, and their potential astrophysical and cosmological applications justify further studies, with particular regard to linearized stability analyses.

The exotic matter concentrated at the throat of a Schwarzschild thin-shell wormhole is described as a two-dimensional perfect fluid featuring negative definite $\sigma$. The determination of EoS for surfaces is difficult even in the case of conventional matter \cite{pcr}. Poisson and Visser \cite{povi} assumed the generic barotropic EoS $p=p(\sigma)$, and analyzed the linear stability of Schwarzschild thin-shell wormholes under perturbations preserving the symmetry. Two oddities of this model arise. First, the squared sound speed in the throat fluid takes values out of the expected range when linearly stable solutions are considered. Poisson and Visser pointed out that discussions of sound speed are problematic if a detailed microphysical model for the thin-shell exotic matter is unavailable. Second, the equilibrium throat radius $a_0=3M$, where $M$ is the wormhole mass, is usually unstable. This feature restricts the size and location of stability regions for common choices of barotropic EoS. Interestingly, the throat radius $a_0=3M$ is also unstable in the thin-shell wormhole model with non-zero cosmological constant proposed by Lobo and Crawford \cite{locra}. More generally, throat radii with distinguished stability properties emerge in $d$-dimensional formulations of thin-shell wormholes including cosmological constant, charge, geometric-topological factor, and barotropic EoS \cite{dile}. Certainly, the occurrence of instabilities narrow the applicability of thin-shell wormholes models with barotropic EoS.

The stability of thin-shell wormholes when $a_0$ nears the Schwarzschild radius $2M$ remains essentially unexplored. GLV have applied their novel stability analysis to asymmetrical Schwarzschild wormholes. This approach deals with external forces acting on the throat, dispenses with the use of EoS linking surface pressure with surface energy density, and imposes constraints on the mass function
\begin{equation}\label{msa}
m_s(a)=4\pi\sigma(a)a^2,
\end{equation}
where $a$ is dynamical throat radius. For the sake of concreteness, the reader is referred to the discussion of Fig. 4 in \cite{glv}, with regard to the particular situation $M_{-}/M_{+}=1, x\rightarrow 1$ i.e. symmetric Schwarzschild thin-shell wormhole with equilibrium radius arbitrarily close to the Schwarzschild radius. These authors observed that the size of the stability regions decreases in this limit. The present author interprets this as a warning, not a diagnosis of instability. In fact, the possible occurrence of stability in this limit for specific choices of $m_s(a)$ was not ruled out by GLV.

Khaybullina et al. \cite{kamsi} applied the cut-and-paste procedure to Schwarzschild black hole and Ellis wormhole solutions. The gluing of these metrics led to highly asymmetric wormhole models featuring thin-shell throats. The stability of these throats was analyzed with the GLV method. These authors showed the increasing difficulty to stabilize their models as $a_0$ nears the Schwarzschild radius, and discussed the significance of the ratio of the two masses involved. The possibility of constructing a particular Schwarzschild-Ellis wormhole which satisfies GLV's inequalities in this limit remains open.

The issues of the Schwarzschild thin-shell wormhole go beyond the instability at $a_0=3M$ and the challenge to stabilize the throat as $a_0\rightarrow 2M$. Eiroa and Simeone \cite{eisi} introduced a thin-shell throat to simplify a truncated Morris-Thorne wormhole model proposed by Lobo \cite{lobo}. Assuming that $p$ and $\sigma$ are related by the cosmologically motivated Chaplygin EoS \cite{chap}, and using methods of dynamical systems, Eiroa and Simeone determined the instability of this model at every equilibrium throat radius under small perturbations which preserve the symmetry \cite{bbc}. Also, Kuhfittig \cite{kuh} solved the energy conservation equation for spherically-symmetric thin-shell wormholes with phantom-like EoS, and considered slight perturbations of the thin-shell potential. This model turned out to be unstable in the whole interval $(2M,+\infty)$.

Provided the fundamental importance of the Schwarzschild thin-shell wormhole model, in this work we propose the use of variable EoS to enhance its stability properties. The viability of variable EoS of the type $p=p(\sigma,a)$ was discussed in \cite{glv}. We generalize Poisson and Visser's analysis to deal with variable EoS, and see how the instability at $a_0=3M$ can be removed. The question arises whether variable EoS can be cosmologically motivated as well.  Variable EoS used in cosmology relate volumetric pressure to volumetric energy density and cosmic scale factor. (See, for example, \cite{gbp,zwz,gz,ud,pdl}). However, the different natures of Schwarzschild thin-shell wormholes and FRW cosmological models preclude straightforward transitions from EoS depending on cosmic scale factor to EoS depending on throat radius. On the other hand, the present author is not aware of any restriction on the possible forms of variable thin-shell EoS imposed by cosmological models \cite{mlc}.

We discuss three types of variable, throat fluid EoS. Our choices are guided by simplicity.
Extending Kuhfittig's approach to linearized stability, we employ these EoS to illustrate the elimination of the instability at $a_0=3M$. In one case we find unobstructed, semi-infinite stability regions with movable boundary. In the other two cases the Schwarzschild thin-shell wormhole is stabilized at every $a_0\in(2M,+\infty)$. Particularly, these results entail linear stability in the limit $a_0\rightarrow 2M$. The stabilization of one of the models is also verified using GLV's restriction on $m_s^{\prime\prime}(a_0)$. Moreover, we linearize the (second-order) equation of motion of the model with variable Chaplygin-like EoS, and link the squared proper angular frequency of radial throat oscillations with the second derivative of the thin-shell potential. Notably, in some cases this quantity remains positive and bounded when the equilibrium radius gets arbitrarily close to the Schwarzschild radius.

To the best knowledge of the present author, the generalization of Poisson and Visser's formula to variable EoS, the elimination of the instability at $a_0=3M$, the full linear stabilization of the Schwarzschild thin-shell wormhole with variable EoS, the confirmation of stability in the limit $a_0\rightarrow 2M$ using GLV's approach, and the boundedness of the throat oscillation frequency in the same limit have not been previously reported in the literature of thin-shell wormholes.

In Section 2 we review Visser's cut-and-paste method for the construction of Schwarzschild thin-shell wormholes. The generalization of Poisson and Visser's linear stability analysis to variable EoS, as well as the possible elimination of the instability at $a_0=3M$ are considered in Section 3. In Section 4 we use Kuhfittig's approach to discuss stability properties of three thin-shell wormhole models featuring variable EoS. Section 5 is devoted to the analysis of small oscillations about arbitrary equilibrium radius for models with variable Chaplygin EoS. Finally, in Section 6 we summarize our results and suggest further approaches to thin-shell wormholes models with variable EoS.

\section{Schwarzschild thin-shell wormholes}
The basic ingredient for the construction of the thin-shell wormhole is the Schwarzschild metric
\begin{equation}\label{sch}
  ds^2=-\left(1-\frac{2M}{r}\right)dt^2+\left(1-\frac{2M}{r}\right)^{-1}dr^2+r^2(d\theta^2+\sin^2\theta\,d\phi^2)
\end{equation}
describing a black-hole spacetime. We take two copies $(\pm)$ of this manifold and remove from each the corresponding four-dimensional region $\Omega^{\pm}=\{r\le a\,|\,a>2M \}$. Next, we identify the time-like hypersurfaces $\partial \Omega^{\pm}=\{r=a\,|\,a>2M \}$ and obtain the only one hypersurface $\partial \Omega$. The arising spacetime manifold is geodesically complete \cite{visser}. It includes two asymptotically flat regions connected through $\partial \Omega$, which constitutes the throat of the wormhole.

Israel's thin shell formalism requires the same induced metric on each side of $\partial \Omega$. Also, it
relates the discontinuity in the extrinsic curvature at the throat to its intrinsic energy-momentum tensor, which describes a two-dimensional perfect fluid.

The derivation of the equations of motion of spherically-symmetric, time-like thin shells with proper time $\tau$, radius $a(\tau)$, surface energy density $\sigma(\tau)$, and surface pressure $p(\tau)$ is a standard application of Israel's formalism. The reader is referred to \cite{epoi} for a pedagogical presentation. A general discussion of thin-shell wormhole construction methods is presented in \cite{glv}.

We choose units $G=c=1$. Einstein's equations for the wormhole take the form
\begin{equation}
\sigma=-\frac{1}{2\pi a}\;\sqrt{1-2M/a+\dot{a}^2}, \label{stswh}
\end{equation}
\begin{equation}
p=\frac{1}{4\pi a}\;\frac{1-M/a+\dot{a}^2+a\ddot{a}}
{\sqrt{1-2M/a+\dot{a}^2}}, \label{ptswh}	
\end{equation}
where overdots denote derivatives with respect to $\tau$.

Equilibrium is defined by $\dot{a}=\ddot{a}=0$. The corresponding surface energy density and surface pressure are given by
\begin{equation}\label{stats}
  \sigma_{0}=-\frac{1}{2\pi a_0}\;\sqrt{1-2M/a_0},
\end{equation}
\begin{equation}\label{statp}
  p_0=\frac{1}{4\pi a_0}\;\frac{1-M/a_0}{\sqrt{1-2M/a_0}},
\end{equation}
where $a_0$ is the throat's equilibrium radius. We see that $\sigma_{0} \rightarrow 0$ and $p_0\rightarrow +\infty$ as $a_0\rightarrow 2M$.  Physically meaningful equilibrium radii satisfy $a_0>2M$.

Equations (\ref{stswh}) and (\ref{ptswh}) entail the energy conservation law
\begin{equation}
\dot{\sigma}+\frac{2}{a}(\sigma+p)\dot{a}=0. \label{dlvcons11}
\end{equation}
We seek solutions of the form $\sigma=\sigma(a)$ for $p=p(\sigma,a)$. In this case we get
\begin{equation}\label{conegen}
  \sigma^{\prime}+\frac{2}{a}\left[\sigma+p(\sigma,a)\right]=0,
\end{equation}
where prime denotes differentiation with respect to $a$.

Equation (\ref{stswh}) can be recast as
\begin{equation}
\dot{a}^2+V(a)=0, \label{emn}
\end{equation}
where the thin-shell potential is given by
\begin{equation}
V(a)=1-2M/a-\left[2\pi a \sigma(a)\right]^2.
\label{epotwhn}
\end{equation}
Each solution of (\ref{conegen}) determines a specific form of $V(a)$.

\section{Linearized stability analysis}
Equation (\ref{emn}) implies that $V(a)$ is not the type of external potential
usually found in classical mechanics. The reason being that the "total energy"
vanishes identically. Accordingly, every perturbation of the kinetic energy term in (\ref{emn})
will be compensated with a perturbation of the potential.
Particularly, if the thin shell is in equilibrium at $a=a_0$ when the
perturbation occurs, the perturbed value of $V(a_0)$ is necessarily negative.

Solutions of (\ref{conegen}) depend on one integration constant. Assuming that the throat is in equilibrium at $a=a_0$, this constant is chosen such that $\sigma(a_0)$ equals the energy density given by (\ref{stats}). As a consequence, each throat in equilibrium is characterized by a specific solution $\sigma=\sigma(a)$ which allows the evaluation of $V=V(a)$ before perturbation \cite{local}. The behavior of the potential in some neighborhood of the equilibrium radius is described with the expansion
\begin{equation}\label{taylor}
  V(a)=V(a_0)+V^{\prime}(a_0)(a-a_0)
  +\frac{1}{2}V^{\prime\prime}(a_0)(a-a_0)^2
  +O\left[(a-a_0)^3\right].
\end{equation}
Linearized stability analysis deals only with the first three terms of this series.

As a consequence of (\ref{epotwhn}) and (\ref{stats}) we find
\begin{equation}\label{vcero}
  V(a_0)=0.
\end{equation}
Differentiation of (\ref{epotwhn}) and evaluation at equilibrium using (\ref{conegen}), (\ref{statp}), and (\ref{stats}) yield
\begin{equation}\label{vpcero}
  V^{\prime}(a_0)=0.
\end{equation}
Further differentiation and use of
\begin{equation} \label{charul}
p^{\prime}=\frac{\partial p}{\partial \sigma} \,\sigma^{\prime}+ \frac{\partial p}{\partial a}
\end{equation}
lead to the expression
\begin{equation}\label{genex}
V^{\prime\prime}(a)=-\frac{4M}{a^3}-8\pi^2\left[(\sigma+2p)^2+2\sigma\left(1+2\,\frac{\partial p}{\partial \sigma}\right)\left(\sigma+p\right)\right]+16\pi^2a\sigma\,\frac{\partial p}{\partial a},
\end{equation}
which generalizes equation (26) of Poisson and Visser's paper to $p=p(\sigma,a)$. Using (\ref{stats}) and (\ref{statp}) we particularize the above expression to equilibrium and obtain
\begin{equation}\label{newvpp}
V^{\prime\prime}(a_0)=-\frac{2}{a_{0}^{2}}\left[\frac{2M}{a_0}+\frac{M^2/a_{0}^2}{1-2M/a_0}+
(1+2\beta_0^2)\left(1-\frac{3M}{a_0}\right)\right]+8\pi\sqrt{1-\frac{2M}{a_0}}\,\gamma_0,
\end{equation}
where $\beta_0^2\equiv \frac{\partial p}{\partial \sigma} \big|_{a=a_0}$ and $\gamma_0 \equiv -\frac{\partial p}{\partial a} \big|_{a=a_0}$.

Equation (\ref{newvpp}) shows that the second derivative $V^{\prime\prime}(a_0)$ is generally non-zero and depends on the chosen EoS through the parameters $\beta_0^2$ and $\gamma_0$.

The above discussion leads to the truncated expansion
\begin{equation}\label{truncv}
  V(a)=\frac{1}{2}V^{\prime\prime}(a_0)(a-a_0)^2,
\end{equation}
where $V^{\prime\prime}(a_0)$ is given by (\ref{newvpp}). We are interested in equilibrium configurations satisfying $V^{\prime\prime}(a_0)>0$. In this case $V(a)$ is approximately described as a convex parabola in a sufficiently small neighborhood of $a=a_0$.

We have seen that every spherically symmetric perturbation of equilibrium leads to a negative value of the perturbed potential at $a=a_0$. Assuming that the shape of $V(a)$ is slightly deformed when the perturbation is very small, we expect the perturbed potential to be definite negative in some neighborhood of $a=a_0$. Furthermore, if the perturbation is sufficiently small, both the sign of $V^{\prime\prime}(a_0)$ and the approximately parabolic shape of the potential remain unchanged after perturbation. Under these conditions, the deformed potential gets two separate zeros within a sufficiently small neighborhood of $a_0$. If these zeros are located at $a=a_1$ and $a=a_2$ ($a_1<a_2$), the perturbed potential is negative definite in the interval $(a_1,a_2)$. Therefore the assumption  $V^{\prime\prime}(a_0)>0$ entails oscillations between
the slightly separated turning points $a_1$ and $a_2$ when the perturbation is sufficiently small. This situation describes linearized thin shell stability against radial perturbations. On the contrary, the case $V^{\prime\prime}(a_0)<0$ implies instability. The present analysis is not conclusive when $V^{\prime\prime}(a_0)=0$ \cite{dmu}.

Let us consider the class of Schwarzschild thin-shell wormhole models characterized by finite or zero $\beta_0^2$ at $a_0=3M$. Under this condition (\ref{newvpp}) yields
\begin{equation}\label{nvpp3m}
  V^{\prime\prime}(3M)=-\frac{2}{9M^2}+\frac{8\pi}{\sqrt{3}}\,\gamma_0,
\end{equation}
where $\gamma_0$ is evaluated at this equilibrium radius. The assumption of generic barotropic EoS implies $\gamma_0\equiv 0$, which entails the negative value $V^{\prime\prime}(3M)=-2/9M^2$. On the other hand, positive values of $\gamma_0$ arising in models with variable EoS could lead to stability at $a_0=3M$.

Lobo and Crawford \cite{locra} analyzed thin-shell wormholes with a non-vanishing cosmological constant ($\Lambda$). These authors used the generic, barotropic EoS which led to their equation (35) for $V^{\prime\prime}(a_0)$. Assuming the EoS is regular at $a_0=3M$ -so that $\eta_0\equiv dp/d\sigma|_{a_0}$ is finite or zero at that point- and that $9\Lambda M^2\neq 1$, the result $V^{\prime\prime}(3M)=-2/9M^2$ follows. Hence the equilibrium radius $a_0=3M$ is unstable in this case as well. Lobo and Crawford's paper does not extend the calculation of $V^{\prime\prime}(a_0)$ to variable EoS. We expect that the derivation of a new formula for $V^{\prime\prime}(a_0)$ based on $\Lambda\neq 0$ and variable EoS would allow the elimination of the instability at $a_0=3M$.

\section{Variable EoS}
We are aware of the total instability of Schwarzschild thin-shell wormhole models based on phantom-like or Chaplygin EoS. In this Section we explore two generalizations of these EoS characterized by explicit, regular dependence on throat radius $a$. Also, motivated by the behavior of $\sigma_0$ and $p_0$ as $a_0\rightarrow 2M$, we consider a variable EoS with linear dependence on $\sigma$ and singular dependence on $a$.

The analyses presented in this Section use Kuhfittig's approach to linearized stability, which involves the following steps:
\begin{enumerate}
  \item Solution of (\ref{conegen}) to obtain $\sigma=\sigma(a)$ for a generic equilibrium configuration;
  \item Use of (\ref{epotwhn}) to get the explicit form of $V(a)$;
  \item Verification of condition (\ref{vcero});
  \item Use of condition (\ref{vpcero}) to constrain the EoS at equilibrium radius $a_0$;
  \item Evaluation of $V^{\prime\prime}(a)$ at every $a_0\in (2M,+\infty)$ to determine stability regions.
\end{enumerate}

We begin with the EoS
\begin{equation}\label{exom}
  p=\frac{A}{a^n}\,\sigma,
\end{equation}
where $A$, $n$ are constants. The case $n=0$ corresponds to the phantom-like EoS discussed in \cite{kuh}.

Plugging (\ref{exom}) into (\ref{conegen}) we obtain
\begin{equation}\label{odes2}
  \frac{d}{da}\left(\sigma^{2}\right)+\frac{4}{a}\left(1+\frac{A}{a^{n}}\right)\sigma^{2}=0.
\end{equation}
This equation admits the solution
\begin{equation}\label{s2sol}
  \sigma(a)^2=
  \sigma_{0}^2\left(\frac{a_0}{a}\right)^{4}
  \exp\!\left[\frac{4A}{n}\left(\frac{1}{{a}^{n}}-\frac{1}{{a_0}^{n}}\right)\right],
\end{equation}
which satisfies the condition $\sigma(a_0)^2=\sigma_0^{2}$.

Using (\ref{s2sol}) and (\ref{epotwhn}) we obtain
\begin{equation}
  V(a)=1-2M/a-4\pi^2 a^2
  \sigma_{0}^2\left(\frac{a_0}{a}\right)^{4}
  \exp\!\left[\frac{4A}{n}\left(\frac{1}{{a}^{n}}-\frac{1}{{a_0}^{n}}\right)\right],
  \label{vvar1}
\end{equation}
which, combined with (\ref{stats}), features the required property $V(a_0)=0$.

Differentiating $V(a)$ with respect to $a$, evaluating the resulting expression at $a=a_0$, and using (\ref{stats}) we find that $V^{\prime}(a_0)=0$ if and only if
\begin{equation}\label{Dsol}
  A=\frac{1}{2}\frac{a_0^{n}M-a_0^{n+1}}{a_0-2M}.
\end{equation}

Further differentiation of $V(a)$, use of the above expression for $A$, and evaluation at $a=a_0$ lead to the result
\begin{equation} \label{vpp1}
V^{\prime\prime}(a_0)=\frac{4n{M}^{2}-\left( 6n+2\right) M a_0+2n{a_0}^{2}}{a_0^3\left({a_0}-2M\right)},
\end{equation}
that we use to determine stability properties at every equilibrium radius $a_0>2M$.

When $a_0$ gets close to $2M$ we find the behavior
\begin{equation}
V^{\prime\prime}(a_0)\approx -\frac {1}{2M(a_0-2M)}
\end{equation}
which indicates instability for arbitrary $n$. We are interested in situations where $V^{\prime\prime}(a_0)$ turns positive at some $a_0>2M$.

The roots of $V^{\prime\prime}(a_0)=0$ are given by
\begin{equation}\label{rm}
a_{01}=-\frac{\sqrt{{n}^{2}+6n+1}-3n-1 }{2n}\,M,
\end{equation}
\begin{equation}\label{rp}
a_{02}=\frac{\sqrt{{n}^{2}+6n+1}+3n+1}{2n}\,M.
\end{equation}
Evaluating these expressions for real $n$ we find that $a_{02}$ takes values in the range $(2M,+\infty)$ only if $n>0$, while $a_{01}$ takes values out of the same interval for every $n$. Therefore equation (\ref{rp}) determines the only one zero of $V^{\prime\prime}(a_0)$ for each $n>0$.

Each value of $a_{02}\in(2M,+\infty)$ defines the boundary of a semi-infinite stability region satisfying $a_0>a_{02}$. This is compatible with the behavior of $V^{\prime\prime}(a_0)$ for $a_0>>2M$, namely
\begin{equation}\label{avpp1}
V^{\prime\prime}(a_0)\approx \frac{2\,n}{{a_0}^{2}},
\end{equation}
which is positive for $n>0$. On the other hand, equilibrium radii $a_0<a_{02}$ are necessarily unstable.

Now we consider the dependence of $a_{02}$ on $n$. For sufficiently small, positive values of $n$ (\ref{rp}) implies
\begin{equation}\label{sn1}
a_{02}\approx \frac{M}{n},
\end{equation}
while very large, positive values of $n$ lead to
\begin{equation}\label{sn2}
a_{02}\approx \left(1+\frac{1}{4n}\right)2M.
\end{equation}
Hence the boundary of the stability region moves away indefinitely from the Schwarzschild radius as $n\rightarrow 0^{+}$, and approaches the would-be horizon as $n\rightarrow +\infty$. Further analysis of (\ref{rp}) shows that $a_{02}/2M$ decreases monotonically when $n$ increases.

We highlight that the equilibrium throat radius $a_0=3M$ becomes linearly stable for some $n$ values. In fact, evaluating
(\ref{vpp1}) at this radius we get
\begin{equation}\label{vpp3M1}
V^{\prime\prime}(3M)=\frac{4n-6}{27{M}^{2}},
\end{equation}
which is positive for $n>3/2$.

The present analysis shows that the stabilization of the Schwarzschild thin-shell model characterized by (\ref{exom}) is partially achieved. The model is necessarily unstable for $a_0$ in the immediate vicinity of $2M$ for every finite, positive value of $n$.

Now we turn our attention to the EoS
\begin{equation}\label{varchap}
p=\frac{1}{a^n}\,\frac{B}{\sigma},
\end{equation}
where $B$ and $n$ are constants. It reduces to the Chaplygin EoS studied in \cite{eisi} when $n=0$.

Following Kuhfittig's approach, we combine (\ref{varchap}) with (\ref{conegen}), and integrate the arising differential equation for the squared surface energy density. Assuming $n\neq 4$ we obtain
\begin{equation}\label{s2chap}
\sigma(a)^2=\frac{4\left(a_0^na^4-a_0^4a^n\right)B+(n-4)a_0^{n+4}\sigma_0^{2} a^n}{(n-4)a_0^{n}a^{n+4}},
\end{equation}
which satisfies the condition $\sigma(a_0)^2=\sigma_0^2$.
Plugging this result into (\ref{epotwhn}) we get the corresponding expression for $V(a)$, which, combined with (\ref{stats}), satisfies (\ref{vcero}). Next we impose condition $(\ref{vpcero})$ and obtain the form of B, namely
\begin{equation}\label{Bchap}
B=\frac{{a_0}^{n-3}M-{a_0}^{n-2}}{8{\pi }^{2}}.
\end{equation}
We use this result to derive, after some algebra, an expression for the second derivative of $V(a)$ at $a=a_0$:
\begin{equation}\label{vppchap}
V^{\prime\prime}(a_0)=\frac{2n-4}{{a_0}^{2}}-\frac{(2n-6)M}{{a_0}^{3}}
\end{equation}

Repeating the above procedure for $n=4$ we get a form of $\sigma(a)^2$ involving logarithms. However, using the corresponding results for $B$ and $V(a)$ we are led to an expression for $V^{\prime\prime}(a_0)$ identical to the case $n=4$ of (\ref{vppchap}). Therefore no separate linear stability analysis is required for this particular choice of $n$.

The thin-shell potential given by (\ref{vppchap}) admits the power series representation
\begin{equation}\label{imv2m}
V^{\prime\prime}(a_0)=\frac{n-1}{4{M}^{2}}-\frac{n+1}{8{M}^{3}}\,( a_0-2M)+\frac{3}{8{M}^{4}}\,{(a_0-2M) }^{2}+...
\end{equation}
when $a_0>2M$ is in the immediate vicinity of $2M$.

From (\ref{vppchap}) we determine that $V^{\prime\prime}(a_0)$ vanishes only at
\begin{equation}\label{rchap}
a_{0r}=\frac{n-3}{n-2}\,M,
\end{equation}
for each real value of $n$. This root takes values in the range $(2M,+\infty)$ only if $n\in(1,2)$.

Two situations emerge if we assume $n\in(1,2)$. First, the zeroth-order truncation of (\ref{imv2m}) assigns finite, positive values to $V^{\prime\prime}(a_0)$ as $a_0$ gets arbitrarily close to $2M$. Second, (\ref{vppchap}) implies negative values of the same parameter for arbitrarily large $a_0$ ($a_0>>2M$). Hence the only root $a_{0r}$ of $V^{\prime\prime}(a_0)$, given by (\ref{rchap}), separates the stability region $a_0<a_{0r}$ from the instability region $a_0>a_{0r}$. Also, $a_{0r}$ increases monotonically and becomes unbounded as $n\rightarrow 2^{-}$, so the model gets fully stabilized in this limit.

In the case $n\leq 1$ negative values of $V^{\prime\prime}(a_0)$ arise in the limits $a_0\rightarrow 2M$ and $a_0\rightarrow +\infty$, as consequences of (\ref{imv2m}) and (\ref{vppchap}), respectively. Also, these two expressions lead to bounded, positive values of $V^{\prime\prime}(a_0)$ in the same limits for $n\geq 2$. Taking into account the absence of roots $a_{0r}\in(2M,+\infty)$ for $n\leq 1$ or $n\geq 2$, we conclude that the thin-shell wormhole model based on (\ref{varchap}) is totally unstable in the first case (which includes the choice $n=0$ i.e. the Chaplygin EoS), and fully stable in the second case.

Interestingly, the evaluation of (\ref{vppchap}) at $a_0=3M$ provides an expression for $V^{\prime\prime}(3M)$ which exactly reproduces (\ref{vpp3M1}). Therefore this particular equilibrium radius also becomes linearly stable for $n>3/2$ in models based on (\ref{varchap}).

Our choices (\ref{exom}) and (\ref{varchap}) have been guided by simplicity. Particularly, these EoS feature regular dependence on (dynamic) throat radius $a$. On the other hand, when we assume a linear relationship between the equilibrium values $p_0$ and $\sigma_0$, respectively given by (\ref{statp}) and (\ref{stats}), we determine a factor of proportionality which becomes unbounded as $a_0\rightarrow 2M$. The appearance of $a_0-2M$ in the denominator of (\ref{Dsol}) is related to this fact. In the case of (\ref{exom}) the singular dependence on $a_0$ is totally contained in $A$. And the explicit dependence of this EoS on $a$ is regular in the same limit.

The unbounded behavior of the ratio $p_0/\sigma_0$ as $a_0\rightarrow 2M$ can also be accounted for by an EoS featuring singular dependence on $a$. In this case the constant factor included in the EoS can be a regular function of $a_0$. More importantly, an EoS featuring explicit, singular dependence on $a$ could have a strong impact on the dynamics of the associated thin-shell wormhole model.

Now we consider the variable, singular EoS
\begin{equation}\label{varsingeos}
p=\frac{Ca}{a-2M}\,\sigma,
\end{equation}
where $C$ is a constant to be evaluated at the equilibrium radius $a_0$. This EoS takes the approximate form $p\approx C\sigma$ for $a>>2M$. Its relationship with the case $n=0$ of $(\ref{exom})$ analyzed in \cite{kuh} is misleading, since the variability of (\ref{varsingeos}) entails substantially different stability properties.

Applying Kuhfittig's method we obtain
\begin{equation}\label{valc}
C=\frac{M-a_0}{2a_0},
\end{equation}
\begin{equation}\label{nsig}
\sigma(a)=-2a_0{\left(\frac{a_0-2M}{a-2M}\right)}^{\frac{M-a_0}{a_0}}\sqrt{1-\frac{2M}{a_0}},
\end{equation}
\begin{equation}\label{psf}
V^{\prime\prime}(a_0)=\frac{2M}{{a_0}^{3}}.
\end{equation}
Restricting the equilibrium radius to the relevant interval $\large(2M,+\infty\large)$, we see that $C$ is a regular function of $a_0$ which approaches the value $-1/2$ as $a_0\rightarrow +\infty$. Also, $V^{\prime\prime}(a_0)$ is definite positive in the same interval. For this reason, the thin-shell wormhole model based on (\ref{varsingeos}) is linearly stable.

We would like to check our assessments of stability with variable EoS against GLV's stability criterion in the absence of external forces. In the case of symmetric Schwarzschild wormholes inequality (83) of \cite{glv} reduces to
\begin{equation}\label{ineq}
a_0 \,m^{\prime\prime}_s(a_0)\ge \frac{2M^2}{a_0^2\left(1-2M/a_0\right)^{3/2}},
\end{equation}
where $m_s(a)$ is defined in (\ref{msa}). Provided that the right side of this expression diverges as $a_0$ approaches $2M$, $m^{\prime\prime}_s(a_0)$ must diverge as well to guarantee stability in this limit.

For the sake of brevity, we focus our attention only on the model with singular EoS. Plugging (\ref{nsig}) into (\ref{msa}), and evaluating $m^{\prime\prime}_s(a_0)$ we obtain
\begin{equation}\label{nineq}
a_0\,m^{\prime\prime}_s(a_0)=\left(\frac{a_0}{M}-1\right)\left[\frac{2M^2}{a_0^2\left(1-2M/a_0\right)^{3/2}}\right].
\end{equation}
Since $a_0>2M$, this result implies that (\ref{ineq}) is fulfilled at every equilibrium radius. Particularly, the wormhole remains stable when $a_0$ gets arbitrarily close to the Schwarzschild radius.

\section{Small oscillations}
The above discussions relate linear stability to the positivity of $V^{\prime\prime}(a_0)$. Apart from its sign, the magnitude of this parameter is usually irrelevant to thin-shell wormhole stability analysis. We have examined two variable EoS which imply bounded values for this second derivative in the whole interval $(2M,+\infty)$. It would be interesting to have a physical interpretation of the magnitude of $V^{\prime\prime}(a_0)$ as well.

The study of oscillations in classical mechanics links the second derivative of the potential energy with squared angular frequency in the case of (exact or approximate) harmonic motion about an equilibrium point. The question arises whether a similar interpretation of $V^{\prime\prime}(a_0)$ could be elaborated, despite the noted differences between thin-shell potentials and classical potential energy functions.

To explore this possible similarity we rewrite (\ref{emn}) in the standard form
\begin{equation}
\frac{1}{2}\dot{a}^2+U(a)=0, \label{memn}
\end{equation}
where the redefined thin-shell potential $U(a)=V(a)/2$ is reminiscent of classical potential energy.

The perturbed values of $U(a)$ are necessarily negative in a sufficiently small neighborhood of equilibrium radius $a_0$ due to the positivity of $\dot{a}^2/2$. Assuming $U(a)\approx U(a_0)+U^{\prime\prime}(a_0)(a-a_0)^2/2$, where $U(a_0)<0$, we see that, after a slight symmetry-preserving perturbation, (\ref{memn}) is replaced with the approximate equality
\begin{equation}\label{clame}
\dot{a}^2+U^{\prime\prime}(a_0)(a-a_0)^2=2|U(a_0)|.
\end{equation}

The equation of motion of a Schwarzschild thin-shell wormhole model emerges when we combine a particular EoS $p=p(\sigma,a)$ with (\ref{ptswh}) and (\ref{stswh}). Let us assume that the equation of motion of certain thin-shell wormhole admits approximate solutions $a(\tau)$ with harmonic $\tau$-dependence, proper angular frequency $\omega_0$, and amplitude $\alpha$. In this case (\ref{clame}) entails
\begin{eqnarray}
  \omega_0^2\alpha^2&=&2|U(a_0)|, \\
   U^{\prime\prime}(a_0)\alpha^2 &=& 2|U(a_0)|,
\end{eqnarray}
which together imply
\begin{equation} \label{uppo2}
\omega_0^2=U^{\prime\prime}(a_0),
\end{equation}
where $U^{\prime\prime}(a_0)$ is necessarily positive.

In virtue of (\ref{uppo2}) the stability criterion  $U^{\prime\prime}(a_0)>0$ is equivalent to $\omega_0^2>0$ in the case of approximate harmonic oscillations. Also, the wormhole throat oscillates in the immediate vicinity of $a=a_0$ with proper frequency $\nu_0=\sqrt{U^{\prime\prime}(a_0)}/2\pi$. These results provide physical meaning to the numerical value of $U^{\prime\prime}(a_0)$ in the case of approximate harmonic motion about equilibrium.

To broaden our scope of the link between $U^{\prime\prime}(a_0)$ and thin-shell throat dynamics, we deal with approximate solutions of the equation of motion associated with the variable Chaplygin EoS, and discuss  $\omega_0^2$ values in all possible cases. From (\ref{varchap}), (\ref{ptswh}), and (\ref{stswh}) we get
\begin{equation}\label{eom}
1-\frac{M}{a}+\dot{a}^2+a\ddot{a}=-\frac{8\pi^2B}{a^{n-2}}.
\end{equation}
An equivalent system of two first-order differential equations was linearized by Eiroa and Simeone \cite{eisi} in the particular case $n=0$. These authors discussed stability properties based on the eigenvalue structure of the associated $2\times2$ matrix. Instead, we analyze the second-order equation (\ref{eom}) directly and find the squared proper angular frequency of the small oscillations for arbitrary real values of $n$.

The dynamical throat radius $a=a(\tau)$ can be written as a perturbation of the equilibrium radius, namely
\begin{equation}\label{critf}
a(\tau)=a_0\left[1+\epsilon(\tau)\right].
\end{equation}
Inserting this expression in (\ref{eom}) we obtain
\begin{equation}\label{epseq}
\frac{1}{a_0^2(1+\epsilon)}-\frac{M}{a_0^3(1+\epsilon)^2}+\frac{\dot{\epsilon}^2}{1+\epsilon}+\ddot{\epsilon}=
-\frac{8\pi^2B}{a_0^n(1+\epsilon)^{n-1}}.
\end{equation}
Assuming $|\epsilon|<<1$, using the binomial approximation, and dropping terms containing $\dot{\epsilon}^2$ we obtain
\begin{equation}\label{lineq}
\ddot{\epsilon}+\omega_0^2\epsilon+\mu=0,
\end{equation}
where
\begin{equation}\label{omega2}
\omega_0^2=\frac{2M}{a_0^3}-\frac{1}{a_0^2}-\frac{8\pi^2(n-1)B}{a_0^n},
\end{equation}
\begin{equation}\label{muexp}
\mu=\frac{1}{a_0^2}-\frac{M}{a_0^3}+\frac{8\pi^2B}{a_0^n}.
\end{equation}
Plugging (\ref{Bchap}) into these expressions we find
\begin{equation}\label{omega2f}
\omega_0^2=\frac{n-2}{a_0^2}-\frac{(n-3)M}{a_0^3},
\end{equation}
\begin{equation}\label{muexpf}
\mu=0.
\end{equation}
We end up with the linearized equation of motion
\begin{equation}\label{oas}
\ddot{\epsilon}+\omega_0^2\epsilon=0
\end{equation}
leading to harmonic oscillations only if $\omega_0^2>0$. Besides, $\omega_0^2<0$ implies instability. In the case $\omega_0^2=0$ a higher order approximation of (\ref{eom}) is required to determine stability properties.

Comparing (\ref{omega2f}) with (\ref{vppchap}), and using the definition of $U(a)$ we arrive at (\ref{uppo2}) again. Thus the validity of this identity is extended to cases with negative or zero $U^{\prime\prime}(a_0)$. We see that stability criteria based on the sign of $U^{\prime\prime}(a_0)$ (or $V^{\prime\prime}(a_0)$), or the sign of $\omega_0^2$ are equivalent. So there is no need to repeat the analysis of (\ref{varchap}) for different values of $n$ based on the sign of $\omega_0^2$.

We emphasize that, as a consequence of (\ref{omega2f}) and (\ref{uppo2}), both $\omega_0^2$ and $U^{\prime\prime}(a_0)$ tend to $(n-1)/8M^2$ as $a_0\rightarrow 2M$. This expression is bounded for finite $n$, and positive whenever $n>1$. Bounded proper angular frequencies are potentially measurable. The above limit provides a link between proper angular frequency, wormhole mass, and adjustable EoS parameter that could play a role in the eventual differentiation of thin-shell wormholes from black holes of the same mass \cite{crude}.

The small oscillations associated with (\ref{varsingeos}) should also be investigated. There are marked differences between the thin-shell potentials derived from (\ref{varchap}) or (\ref{varsingeos}), particularly with regard to the behavior of the third and higher derivatives of $V(a)$ evaluated at $a_0$ in the limit $a_0\rightarrow 2M$. Subtleties may arise in the perturbative analysis of the equation of motion in the case of singular EoS. We leave the corresponding calculation for future research.

\section{Conclusion}
Most of the previous approaches to thin-shell wormhole stability have assumed barotropic EoS for the throat fluid. As a result the equilibrium throat radius $a_0=3M$ is usually unstable. The phantom-like EoS as well as the Chaplygin EoS lead to totally unstable models. The modified Chaplygin EoS entails stability regions, but leaves the instability at $a_0=3M$ untouched due to its barotropic character.

The fundamental role and relative simplicity of the Schwarzschild thin-shell wormhole solution have motivated our linear stabilization procedures based on variable EoS.

Specific choices of variable EoS are less obvious than the usual, cosmologically-motivated selection of phantom-like, Chaplygin, or modified Chaplygin (barotropic) EoS. Although some examples of variable cosmological EoS are available, their implications on the choice of variable, thin-shell wormhole EoS depending on throat radius are uncertain.

We have extended Poisson and Visser's formula for $V^{\prime\prime}(a_0)$ to determine possible effects of variable EoS on linear stability. The procedure is quite general. It does not require specific choices of variable EoS, or involves the integration of the energy conservation equation. And it shows the possible elimination of the instability at $a_0=3M$ for suitable choices of variable EoS.

Our modifications of the the phantom-like and Chaplygin EoS have been accomplished through the introduction of variable coefficients in these equations. In two cases the coefficients take the form of arbitrary powers of dynamical throat radius $a$, where the exponents are free parameters. In another case the coefficient is a fixed function of $a$ which becomes unbounded as $a$ approaches the Schwarzschild radius.

We have employed Kuhfittig's approach to linear stability analysis, which involves the solution of the (first-order) energy conservation equation for surface energy density, as well as the determination of the thin-shell potential and its second derivative at every equilibrium radius $a_0>2M$.

The generalized form of the phantom-like EoS considered here leads to a sequence of semi-infinite stability regions with boundaries approaching the Schwarzschild radius as the (positive) free parameter of the EoS becomes unbounded. On the other hand, for every finite, positive value of this parameter the thin-shell throat is unstable in the immediate vicinity of the would-be horizon. The proposed generalization of the Chaplygin EoS entails models which are totally unstable, partially stable, or fully stable depending on the value of the free parameter. The second modification of the phantom-like EoS, characterized by singular $a$-dependence, entails fully stabilized models.

Using the variable Chaplygin EoS, as well as the singular form of the phantom-like EoS, we have determined linear stability with bounded $V^{\prime\prime}(a_0)$ at every $a_0>2M$. The full linear stabilization of the model with singular EoS has been also confirmed using GLV's condition on $m_s^{\prime\prime}(a_0)$.

Complementarily, we have studied the small oscillations of thin-shell throats satisfying the variable Chaplygin EoS. This approach is based on the linearization of the corresponding (second-order) equation of motion. We have found that the squared proper angular frequency $\omega_0^2$ equates the second derivative of the (redefined) thin-shell potential $U^{\prime\prime}(a_0)$. This result shows the equivalence of the small oscillations analysis and Kuhfittig's method when applied to this EoS. It also indicates that $\omega_0^2$ remains positive and bounded in the limit $a_0\rightarrow 2M$ for various choices of the free EoS parameter $n$.

To the best knowledge of the present author the extension of Poisson and Visser's approach to variable EoS; the application of Kuhfittig's stability analysis to Schwarzschild thin-shell wormholes with variable forms of the phantom-like EoS (with regular or singular $a$-dependence) or variable Chaplygin EoS; the proof of stability in the limit $a_0\rightarrow 2M$ based on GLV's inequality; the direct perturbative analysis of the (second-order) equation of motion for the variable Chaplygin EoS showing full compatibility with Kuhfittig's method; and the possibility of (approximate) harmonic throat oscillations with bounded proper angular frequency in the limit $a_0\rightarrow 2M$ are new results in the literature of these solutions. It should be emphasized that the discussion of thin-shell wormhole oscillations -beyond stability assessments- has received relatively little attention by researchers so far.

We have not calculated the speed of sound in exotic throat fluids with variable EoS. The use of variable EoS could mitigate the sound speed inconformity within a macroscopic treatment. An immediate challenge is the derivation of a sound speed formula applicable to throat fluids with variable EoS.

The three types of variable EoS examined in this paper could also be used in stability analyses of thin-shell wormholes constructed with non-vacuum metrics. The existence of local solutions for the energy conservation equation around the equilibrium radius is guaranteed for external force terms depending exclusively on surface energy density and throat radius \cite{glv}. Thin-shell gravastar models can incorporate variable EoS as well. Local stability analyses will be available for these sources whenever the external force term in the energy conservation equation satisfies the same requirement as above \cite{gtsgs}. The use of singular EoS in this context could be interesting for another reason \cite{tsgsm}.

We expect that the eventual extension of Dias and Lemos' work to variable EoS will lead to enhanced stability properties for thin-shell wormholes in $d$-dimensional general relativity.

Interestingly, the connection between Born-Infeld scalar fields and variable Chaplygin EoS has been explored in cosmology \cite{gz}. Specific scalar field models of thin shells have been analyzed in the literature \cite{sfmts}. The question arises whether a prospective Born-Infeld model for throat fluids would entail (\ref{varchap}).

The use of variable EoS could boost the stability of thin-shell models of intra-galactic wormholes based on the Mannheim-Kazanas-de Sitter solution \cite{bola}. Also, it could improve the stability of truncated Morris-Thorne wormhole models involving dynamical thin-shell boundary \cite{kuh2}. Future developments in these directions could offer perspectives on the stability analysis of other types of galactic wormhole models \cite{rah} and their differentiation from black holes \cite{liba}.

\section{Acknowledgments}
The author would like to thank Dr. M. F. A. da Silva for discussions on thin-shell gravastar models during the period 2010-2011. Dr. Orlando Oliveira provided advice on the use of the Maxima computer algebra system. The author is grateful to him. Thanks to Professor Peter Kuhfittig for useful comments on a preliminary version of this work. The author is indebted to Professor Graham Hall and Mr. Harold Aschmann for suggestions regarding the literature of thin-shell wormholes; and owes very special gratitude to Dr. Olga Savasta for her computational assistance.


\begin{thebibliography}{99}
\bibitem{crude}Our expression for $\omega_0^2$ in the limit $a_0\rightarrow 2M$ leads to the formula
$\nu_0=\sqrt{(n-1)/8}c^3/2\pi GM$ for proper frequency (in SI units). To get an idea of the numerical values involved we assume $n=2$ and consider $M$ values in the range $10^5-10^9 M_{\odot}$ corresponding to supermassive blackholes. As a result $\nu_0$ takes values in the range $10^{-5}-10^{-1}$ Hz. Whether these pulsations could have observable effects on radiative phenomena occurring at or near the wormhole throat is an open question. The same puzzle would arise if proper frequencies are determined for thin-shell wormholes constructed with non-vacuum metrics.
\end{thebibliography}
\end{document}